\documentclass[fleqn,usenatbib]{mnras}

\usepackage{newtxtext,newtxmath}

\usepackage[T1]{fontenc}

\DeclareRobustCommand{\VAN}[3]{#2}
\let\VANthebibliography\thebibliography
\def\thebibliography{\DeclareRobustCommand{\VAN}[3]{##3}\VANthebibliography}

\usepackage{graphicx}	
\usepackage{amsmath}	
\usepackage{tablefootnote}
\usepackage{longtable}
\usepackage[dvipsnames]{xcolor}
\usepackage{float}
\usepackage{appendix}

\title[Stellar Mass-Halo Mass Relation]{Photometric Mass Estimation and the Stellar Mass-Halo Mass Relation for Low Mass Galaxies}

\author[Zaritsky \& Behroozi]{Dennis Zaritsky
\thanks{E-mail: dennis.zaritsky@gmail.com} 
and Peter Behroozi 
\\
\\
$^{1}$Steward Observatory and Department of Astronomy, University of Arizona, Tucson, AZ 85721, USA\\
$^{2}$Division of Science, National Astronomical Observatory of Japan, 2-21-1 Osawa, Mitaka, Tokyo 181-8588, Japan
}

\date{Accepted 2022 December 1. Received 2022 November 30; in original form 2022 August 26}

\pubyear{2022}

\begin{document}
\label{firstpage}
\pagerange{\pageref{firstpage}--\pageref{lastpage}}
\maketitle

\begin{abstract}
We present a photometric halo mass estimation technique for local galaxies that enables us to establish the stellar mass-halo mass (SMHM) relation down to stellar masses of 10$^5$ M$_\odot$.
We find no detectable differences among the SMHM relations of four local galaxy clusters or between the cluster and field relations and we find agreement with extrapolations of previous SMHM relations derived using abundance matching approaches. We fit a power law to our empirical SMHM relation and find that for adopted NFW dark matter profiles and for M$_* < 10^9$ M$_\odot$, the halo mass is
M$_h = 10^{10.35\pm0.02}({\rm M}_*/10^8 {\rm M}_\odot)^{0.63\pm0.02}$.
The normalisation of this relation is susceptible to systematic modelling errors that depend on the adopted dark matter potential and the quoted uncertainties refer to the uncertainties in the median relation. For galaxies with M$_* < 10^{9}$ M$_\odot$ that satisfy our selection criteria, the scatter about the fit in $M_h$, including uncertainties arising from our methodology, is 0.3 dex.
Finally, we place lower luminosity Local Group galaxies on the SMHM relationship using the same technique, extending it to M$_* \sim 10^3$ M$_\odot$ and suggest that some of these galaxies show evidence for additional mass interior to the effective radius beyond that provided by the standard dark matter profile. If this mass is in the form of a central black hole, the black hole masses are in the range of intermediate mass black holes, $10^{(5.7\pm0.6)}$ M$_\odot$, which corresponds to masses of a few percent of M$_h$, well above values extrapolated from the relationships describing more massive galaxies.
\end{abstract}




\section{Introduction}
\label{sec:intro}

How do stars populate the dark matter halos that define the galaxy population? A simple, first order answer is provided by the stellar mass-halo mass (SMHM) relation for galaxies. Measuring that relation, however, is not simple.

There are broadly two ways that this measurement is approached for dwarf galaxies. In the first, using forward modeling or statistical arguments an association is made between the population of dark matter halos theoretically expected to inhabit a certain volume of the Universe and the galaxies observed within the same volume. The association is constrained using observables, such as stellar mass, but the question of scatter in the relation and  simultaneously reproducing all known properties of galaxies, such as clustering or lensing as a function of magnitude or colour, complicates that association \citep[for a review see][]{wechsler}. This approach generally does not provide the relation for each individual galaxy, but is able to bring to bear the tremendous statistical power of today's large surveys and simulation volumes. Exceptions include the more focused analysis of the satellite population of the Milky Way \citep{nadler20,manwadkar,chen22}. In the second, measurements of the internal kinematics of each individual galaxy are used to obtain a dynamical estimate of the mass enclosed at the corresponding radius and an extrapolation based on an adopted dark matter halo model provides an estimate of the halo mass. A sample of galaxies for which this can be done is then used to produce the SMHM relation \citep{dutton, read17}. This approach is able to provide the SMHM relationship for individual galaxies, but is statistically limited to smaller samples due to the required kinematic measurements. Although these two approaches are generally applied independently of each other, there are now some examples of joint analyses \citep[e.g.,][]{yasin}. 

A current weakness in the application of either of these approaches is their inability to track the SMHM relation outside the LG significantly below a  halo mass of $10^{10}$ M$_\odot$. The difficulty arises because such galaxies are rare in redshift surveys beyond the local volume and measurements of the internal kinematics are increasingly difficult for fainter, lower surface brightness galaxies. 
This limitation is unfortunate in that a variety of interesting questions, related both to galaxy evolution and the nature of dark matter, would benefit from an understanding of the low mass SMHM relation. 

We address this current weakness using a novel approach to estimate halo masses for a range of galaxy samples in the literature. Our approach follows the kinematic approach in spirit in that we estimate the mass for each galaxy in our sample and build up the SMHM from observations of many such galaxies. However,
we do not use kinematic measurements, but rather  develop a photometric method that enables us to make the mass estimate independent of any measurement of the internal kinematics of each galaxy. As such, we are able to construct the SMHM using many low mass galaxies and extend the SMHM well below current limits in mass using large samples. In \S\ref{sec:fm} we describe our approach to estimating the galaxy's halo mass (baryons + dark matter within an estimated virial radius, see \S\ref{sec:mass}) denoted M$_h$. In \S\ref{sec:results} we present the SMHM relations for low mass cluster and field galaxies separately and jointly and discuss the effect of uncertainties (both observational and theoretical). In \S\ref{sec:discussion} we extend the relation to even lower mass by adding Local Group dwarfs and speculate on the possibility of inferring the masses of intermediate mass black holes in these galaxies. We adopt a WMAP9 flat $\Lambda$CDM cosmology with H$_0 = 69.7$ km s$^{-1}$ Mpc$^{-1}$ and $\Omega_m = 0.281$ \citep{hinshaw} for consistency with some previous studies to which we compare.

\begin{table*}
\centering
\caption{Samples used}
\begin{tabular}{ lccccrrl } \hline  
Source & Type & Band & $v_c$ & $\sigma$ & Sample & with M$_h$ & Notes \\ 
&&&&&Size&\\
\hline
\cite{chilingarian} & dE & B & --- & \checkmark & 46 &\\

\cite{collins} & LG dSph & V & --- & \checkmark & 36 && tidal objects removed\\
\cite{geha} & dE & V & \checkmark & \checkmark & 17 & \\
\cite{jorgensen} & E \& S0 & r & --- & \checkmark & 280 && in clusters\\


\cite{mieske} & UCD &V & --- & \checkmark & 15&& only those they observed\\
assorted (see text) & UDGs & various & --- & \checkmark & 19 & \\
\hline
\cite{blanton} & assorted & r & --- & --- & 49968 &1661& SDSS low z NYU-VAGC \\
\cite{nadler} & LG galaxies & V & --- & --- & 54 & 24& LG galaxy compilation\\
\cite{virgo} & dwarfs & g& --- & ---&404&240& Virgo\\
\cite{lamarca} & dwarfs & r& --- & --- &317&257& Hydra\\
\cite{mao} & satellites & r& --- & --- & 127 &119& SAGA survey\\
\cite{park1,park2} & dwarfs & I& --- & --- &87  &82& NGC 2784 \& NGC 3595 groups \\
\cite{venhola} & dwarfs &r & ---  & ---&564&477& Fornax\\
\cite{yagi} & dwarfs & r & --- & --- &751&685& Coma\\
\end{tabular}
\label{tab:samples}
\end{table*}

\begin{table}
\centering
\caption{Coefficients for Equation \ref{eq:m2l}}
\begin{tabular}{ lrr } \hline  
Coeff. & Optical & Near-IR \\ 
\hline
a&$0.198^{+0.024}_{-0.024}$&0.443$^{0.031}_{-0.030}$\\
b&$0.140^{+0.071}_{-0.072}$&$-0.978^{0.122}_{-0.128}$\\
c&$0.192^{+0.006}_{-0.006}$&$0.158^{0.007}_{-0.007}$\\
d&$-0.923^{+0.019}_{-0.019}$&$-0.967^{0.029}_{-0.031}$\\
e&$-0.108^{+0.019}_{-0.019}$&$-0.040^{0.017}_{-0.016}$\\
f&$1.306^{+0.056}_{-0.056}$&$2.187^{0.102}_{-0.103}$\\
\label{tab:coefficients}
\end{tabular}
\end{table}

\section{Estimating Halo Masses}
\label{sec:fm}

Because we aim to estimate halo masses for as many galaxies as possible, we develop an estimator based solely on photometric properties, bypassing measurements of internal kinematics.
We divide the task of developing a broadly applicable estimator into two steps. In the first, we estimate the enclosed mass within the effective radius, $r_e$. The choice of the effective radius as a standard radius in galaxy photometry has generally been justified as a compromise between a sufficiently small radius where one obtains high signal-to-noise measurements and a sufficiently large one that encloses a large fraction of the luminous mass. However, in our context the choice is particularly fortuitous because a simple enclosed mass estimator at this radius is surprisingly robust to the internal detailed structure of galaxies \citep[see \S\ref{sec:wolf};][]{wolf}.  In the second step, we fit dark matter halo models, constrained to match the enclosed dark matter mass within $r_e$ obtained in the first step, to estimate the halo mass. 

What we propose in the first step is the novel part of our approach. This approach has the potential to increase the number of galaxies with estimated  halo masses by the ratio of the size of photometric to spectroscopic samples, much in the same way that photometric redshifts greatly increase the numbers of galaxies available for study. Again, as with photometric redshifts, one exchanges this gain in sample size for the precision obtained for each individual case and the added potential for the occasional catastrophic error.
The second step in our procedure is not new and has been taken previously using kinematically-constrained enclosed mass estimates at $r_e$, and other specific radii, in a variety of ways by various investigators \citep[e.g.,][]{dutton,read17}. 

\subsection{Step 1: estimating $M_e$}
\label{sec:wolf}

Scaling relations provide an opportunity to take the first of the two steps. By providing relationships among measured parameters, the appropriate scaling relation, plus the assumption that the galaxies of interest lie on the scaling relation, can be used to recover missing data for those galaxies. Historically, examples of this type of approach predominantly focus on the use of scaling relations to estimate distances, as in the use of the relationship between luminosity and rotation velocity for disk galaxies \citep{tully}. Occasionally, those same relations can be used to recover another of the related parameters \citep[e.g., rotation velocity using the Tully-Fisher relation;][]{gonzalez}. The work presented here most closely resembles the \cite{gonzalez} study.

To be able to pull off this trick most broadly, the scaling relation must be applicable to all galaxies, not just to a subset of galaxies such as disk galaxies.
Across several studies, we have developed and applied a universal scaling relations for stellar systems \citep{zar06,zar06a,zar08,zar11,zaritsky12}. In those papers, we showed that galaxies that span the known range of luminosities and morphologies satisfy a relationship between $r_e$, a measure of the internal kinematics (the velocity dispersion, $\sigma$, for pressure supported systems or a combination of rotational velocity and $\sigma$ for systems with significant dynamical support from both), and the projected mean surface brightness within $r_e$, $I_e$. The parameters involved are those also found in the Fundamental Plane scaling relation \citep{djorgovski,dressler}, but the functional form is more complex to allow for the broader range of systems to which it applies. The value of having a scaling relation for all galaxies is that we can apply a methodology based on it without restriction or any prior knowledge of the galaxy to which it is being applied \citep[see][for an example that combines results using Tully-Fisher for spirals and Fundamental Plane for giant spheroids]{dutton}.

To calibrate the derived estimates of the enclosed mass, $M_{e}$, within a sphere of radius $r_e$, we present an alternative approach to that of our previous papers.  We start with the well-established, widely-adopted mass estimator from \cite{wolf}. In that study, \cite{wolf} found, based on simulations, a mass estimator that is robust against changes in the internal spatial and kinematic details of the spheroidal stellar system\footnote{The \cite{wolf} estimator was validated only for spheroidal galaxies, but the empirical scaling relation is valid for both disks and spheroids if the appropriate kinematic measurement is used for disks \citep{zar08}. As such, once the enclosed mass estimates are calibrated for spheroidal galaxies, then the estimates are calibrated for all galaxies.} Their estimator for the mass enclosed within a sphere of radius $r_e$ is
\begin{equation}
    M_{e} = 930\sigma^2 r_{e},
\label{eq:wolf}    
\end{equation}
\noindent
where $\sigma$ is the line of sight velocity dispersion in km sec$^{-1}$, $r_e$ is the effective radius of the surface brightness profile in pc, and $M_e$ is in solar units.
By calibrating our results to the 3-D enclosed mass, we are taking a slightly different approach than our earlier empirical one \citep{zar06,zar08} that worked entirely in projected quantities. As such, the scaling relation presented here has minor quantitative differences from that presented previously.
We repeat for emphasis that the observed quantities ($r_e$ and $I_e$) are projected but that the derived quantity (M$_e$) is not.

Between the two measurements needed to apply this estimator, $\sigma$ is by far the more challenging to obtain, particularly for low luminosity galaxies. As such, it is  particularly advantageous to express the mass estimator exclusively in terms of photometric measurements. To do this, we first define the enclosed 3-D mass as the enclosed projected luminosity times an effective mass-to-light ratio, $\Upsilon_{e}$, to rewrite Eq. \ref{eq:wolf} as
\begin{equation}
    \pi r_e^2 I_e \Upsilon_{e} = 930\sigma^2 r_e,
\label{eq:wolf_rewrite}
\end{equation}
\noindent
where $I_e$ is the mean surface brightness within $r_e$ in units of L$_\odot/{\rm pc}^2$ and $\Upsilon_{e}$ is given in solar units.  Taking the logarithm (all logarithms presented in this paper are base 10) of both sides, expressing $r_e$ in kpc, and organizing terms we find
\begin{equation}
    \log r_e = 2 \log \sigma - \log I_e - \log \Upsilon_{e} - 0.53.
\label{eq:fm}    
\end{equation}
So far, this is simply a different expression of the \cite{wolf} mass estimator.

To eliminate or solve for $\sigma$,  we need a second equation involving the two unknowns, $\sigma$ and $\Upsilon_{e}$.  At this point progress requires an ansatz for the functional form of $\Upsilon_{e}$. A natural (i.e., simple) proposal is that $\log \Upsilon_{e} = f(\log\sigma)$.

\begin{figure}
\centering
\includegraphics[scale=0.3]{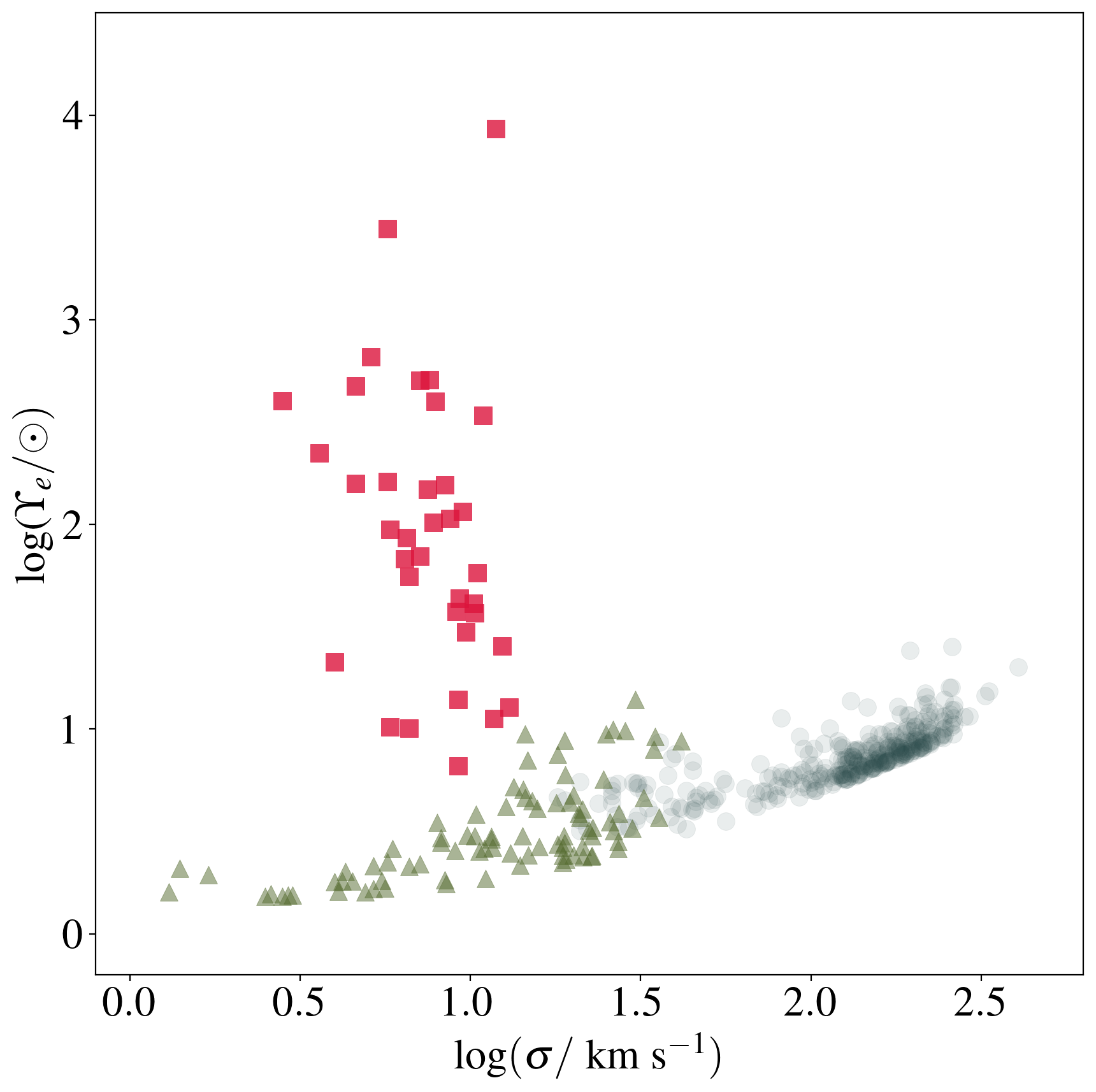}
\caption{The effective mass-to-light ratio, $\Upsilon_{e}$, such that $M_{e} = \pi r_e^2 I_e \Upsilon_{e}$, vs. $\sigma$. Three branches can be distinguished. Toward larger $\sigma$ the space is that populated by ellipticals and dwarf ellipticals (lightly coloured circles). At smaller $\sigma$ there are two branches, that populated by dsph and ultrafaint galaxies satellites of the Milky Way and M31 (red squares) and that populated by compact dwarf galaxies (green triangles).}
\label{fig:m2l}
\end{figure}

To guide our understanding of what form such a function might take we evaluate $\Upsilon_{e}$ using Equation \ref{eq:fm} and plot those values vs. $\sigma$ for a wide range of spheroidal stellar systems with spectroscopically measured $\sigma$'s \citep[see Table 1;][]{jorgensen, geha,mieske,collins,chilingarian}
in Figure \ref{fig:m2l}. 
We have made one set of edits to the literature sample in that we have removed five galaxies from the \cite{collins} sample of LG dwarfs that are suspected to be experiencing significant tidal forces (Crater II \citep{sanders}; Wilman I and Triangulum II \citep{fritz}; Hercules I \citep{fu}); and Leo V \citep{collins17}). While there is a dependence of  $\Upsilon_{e}$ on  $\sigma$, there is also a bifurcation in behaviour at low  $\sigma$. The two branches highlight the divergence in properties between high and low surface brightness stellar systems. Given this behaviour, it is manifestly not possible to describe $\Upsilon_{e}$ as only a function of $\sigma$. We conclude that any appropriate functional form must at least also include $I_e$.

The next simplest ansatz is that 
$\log \Upsilon_{e} = f(\log\sigma,\log I_e)$
and that this function is first order in both $\log\sigma$ and $\log I_e$. Such a proposition leads to equations of the form of the Fundamental Plane \citep{dressler,djorgovski}, which has been so successful at describing giant ellipticals, but which fails to describe low luminosity spheroids. The cause of that failure is also evident in Figure \ref{fig:m2l}. One can only describe $\log \Upsilon_{e}$ adequately with a linear function of log $\sigma$ for the higher $\sigma$ stellar systems.

The next step in complexity is adopting a function $f$ that is second order in $\log \sigma$ and $\log I_e$,

\begin{equation}
\begin{split}
\log \Upsilon_{e}  &=   a\ ({\log {\sigma}})^2 +  b\ {\log {\sigma}} + c\ (\log I_e)^2  +  \\
& \hspace{50pt} d\ {\log {\rm I_e}} + e\ {\log I_e}\log \sigma + f,
\end{split}
\label{eq:m2l}
\end{equation}
where we neglect cross terms that are leading second order but discuss them further below.

We evaluate the coefficients in Eq. \ref{eq:m2l} by replacing  $\log \Upsilon_{e}$ in Eq. \ref{eq:fm} with the right hand side of Eq. \ref{eq:m2l} and fitting the data shown in Figure \ref{fig:m2l} plus a compilation of ultra-diffuse galaxies  
\citep{vdk17,vdk19,toloba,chilingarian,martin-navarro,gannon} to extend further the range of galaxy types. We do make one further edit of the literature data in that we exclude systems with $r_e < 10$ pc, which are predominantly globular clusters but do include some ultra-compact dwarfs. There are not many such systems in the sample, so the derived coefficient values are not significantly affected by this choice, but there are indications that these systems start to deviate from a scaling relation of this form \citep[][and this work]{forbes08}. We believe this deviation happens because such compact systems are completely stellar dominated within $r_e$, and therefore have an $\Upsilon_{e}$ that is independent of $I_e$ and $\sigma$, making it difficult for a low order functional form to adequately adjust to such behaviour. 

As expected from such a large list of disparate studies, the data are a heterogeneous set of photometric and kinematic measurements. We place the  surface brightnesses on a comparable system of solar luminosities, appropriate for each band \citep{willmer}, but make no correction for color differences between the galaxies and the Sun. There is also no correction for how color gradients might affect $r_e$ or how kinematic measurements vary between central values of $\sigma$ and aperture values. All of these irregularities among data sets and improper or ignored corrections can be expected to lead to less rather than more coherence in the resulting scaling relation. Our eventual estimation of the precision of our mass estimates using this set of data is therefore an upper limit on the intrinsic scatter.

\begin{figure}
\centering
\includegraphics[scale=0.2]{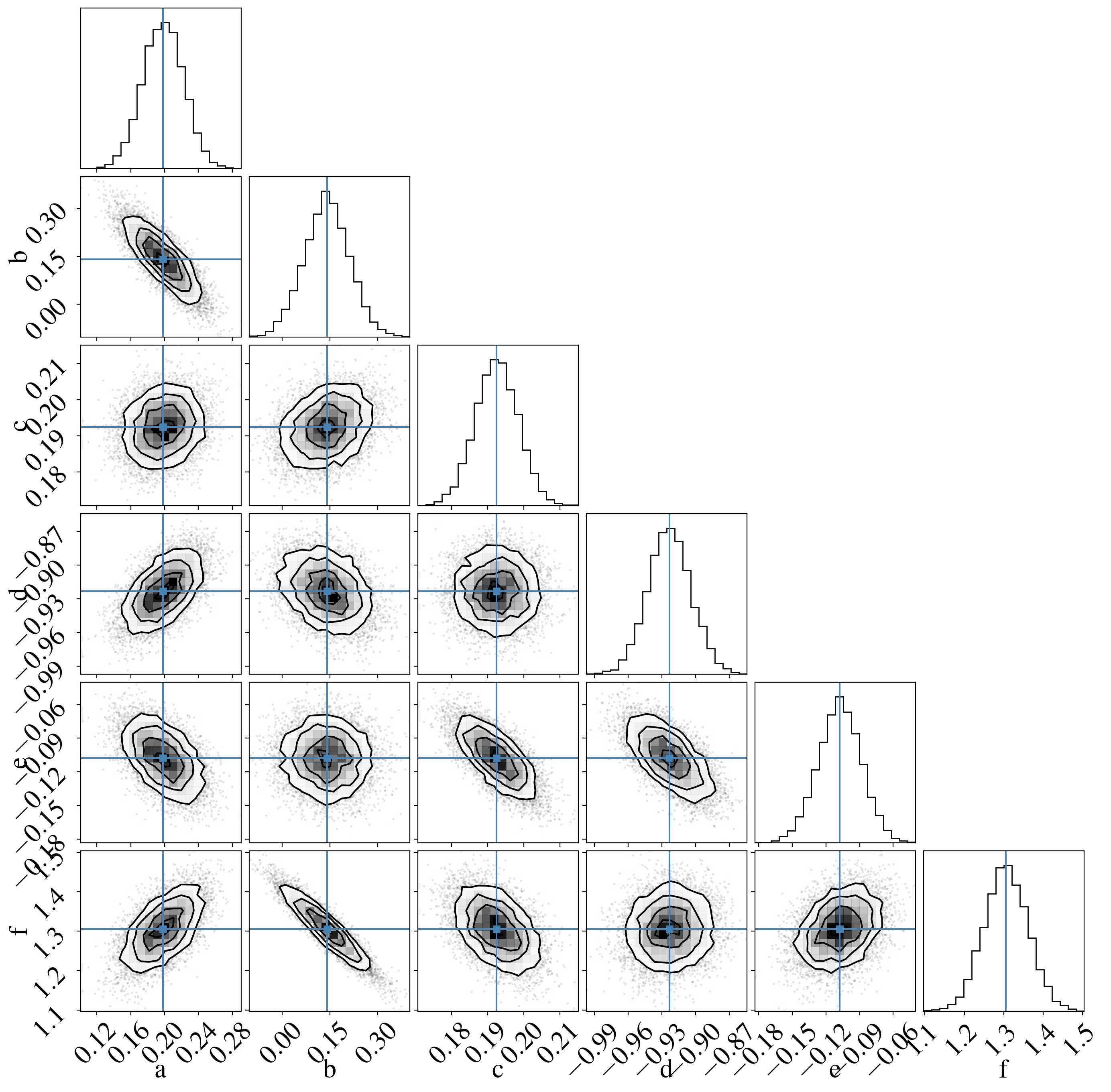}
\caption{Posterior distributions of each of the coefficiencts in Equation \ref{eq:m2l}.}
\label{fig:corner}
\end{figure}

\begin{figure*}
\centering
\includegraphics[scale=0.45]{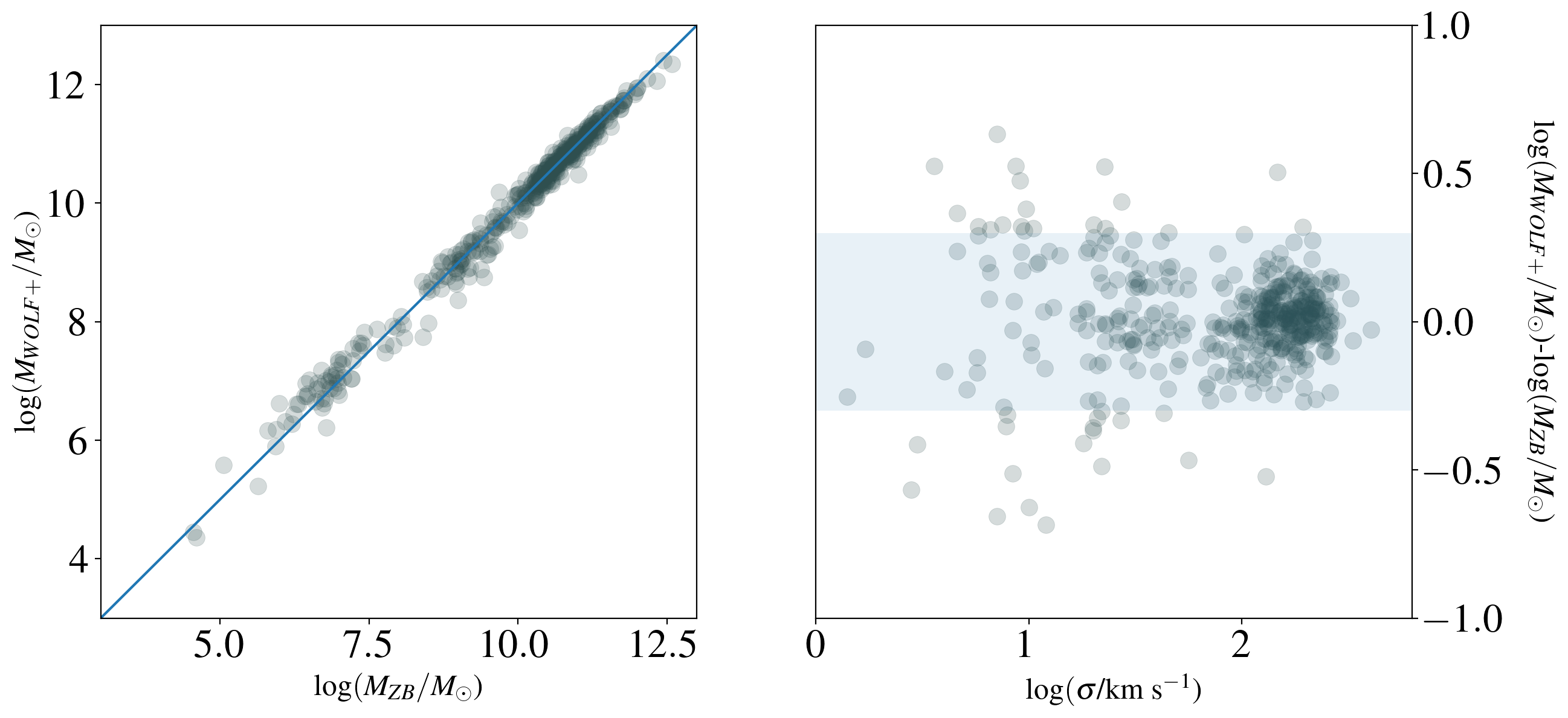}
\caption{Comparison of the inferred enclosed masses at $r_e$ using the \citet{wolf} estimator, M$_{WOLF+}$, and our scaling-relation based estimate, M$_{ZB}$, for the optical galaxy sample (see Table \ref{tab:samples}). Left panel shows the derived values and the 1:1 line and includes both the high and low surface branches visible in Figure \ref{fig:m2l}. Right panel shows the differences in the two estimates vs. $\sigma$. The shaded region encloses values that are within a factor of two of each other.}
\label{fig:wolf_test}
\end{figure*}

We derive the coefficients using a Bayesian approach and the {\sl emcee} Python implementation of a Markov chain Monte Carlo sampler \citep{emcee}. The model is assumed to have no intrinsic scatter and be as given by Equation \ref{eq:m2l}. We adopt uniform priors on all of the parameters and parameter ranges that avoid resulting posterior distributions that peak near the range edges. The
corner plot showing the character of the uncertainties in the coefficients is presented in Figure \ref{fig:corner} and the resulting coefficient values are listed in Table \ref{tab:coefficients}. The correspondence between our estimate of M$_e$ and that obtained using the spectroscopically-measured $\sigma$ and the \cite{wolf} estimator is excellent (Figure \ref{fig:wolf_test}), with a standard deviation about the 1:1 line of 0.17 dex (corresponding to a relative error of $\sim$ 50\%). In the right panel of Figure \ref{fig:wolf_test} we show that the majority of the estimates are within a factor of two of the \cite{wolf} values, with larger scatter for systems with $\sigma \lesssim$ 10 km s$^{-1}$ although at these low values of $\sigma$ there are large fractional uncertainties in the spectroscopically measured $\sigma$'s as well. As such, we cannot ascertain whether the larger scatter is due to intrinsically larger scatter about our scaling relation or observational errors in the spectroscopically-determined values of $\sigma$. In either case, there is no evident systematic residual with $\sigma$ although one must remain aware that binaries in these lowest mass galaxies could lead to an upward bias in the measured $\sigma$'s and hence in the functional form of the fit as well.

To further explore the nature of the scatter, we now redo the analysis with a sample of K-band photometry for spheroidal stellar systems from \cite{forbes08}. As described by those authors, the advantage provided by using near-IR is a decreased sensitivity to variations in the stellar mass-to-light ratios. For our purposes, we also benefit from the single-source nature of the photometry and analysis. 
The result of applying the same procedure to these data, which includes a similarly diverse range of stellar systems, is a scatter about the 1:1 relation between the \cite{wolf} estimator and ours of 0.17 dex, exactly what the optical estimates yielded\footnote{There are two significant outliers from the 1:1 relation. Excluding these two yields a standard deviation of 0.16 dex, still nearly identical to the optical results.}. We conclude that the use of a wide variety of studies in the optical did not contribute significantly to the scatter in our mass estimates. We favor the use of the optical relation going forward because there is so much more data currently available that we can use in our subsequent analysis.

Finally, returning to the choice we made to neglect the cross terms that are leading second order terms in Equation \ref{eq:m2l}, we redo the coefficient fitting including those terms and find that both of the resulting coefficients, for the $\log\sigma (\log I_e)^2$ and $(\log \sigma)^2 \log I_e$ terms, are  consistent with zero. Of course, neither higher order functions or the inclusion of other parameters are excluded by our analysis, but a function of the form presented in Equation \ref{eq:m2l} appears to be the simplest that can adequately express the behaviour seen in Figure \ref{fig:m2l} to the current level of observational precision.

\subsection{Step 2: extrapolating from M$_{e}$ to M$_h$}
\label{sec:mass}

The use of the scaling relation only provides an estimate of the mass interior to $r_e$. To calculate the halo mass, M$_h$, we need an estimate of the dark matter mass. To obtain this estimate, we subtract the contribution to M$_e$ from stars projected within $r_e$ and then determine the parameters of an NFW dark matter density profile \citep{navarro} that best reproduces the remaining mass, the dark matter, within a sphere of radius $r_e$, or M$_{e,DM}$. 
To estimate the stellar mass within $r_e$, we convert the luminosity within $r_e$ to stellar mass by adopting stellar $M/L$ ratios that are either color-dependent \citep{roediger} when a color is available or fixed in the case where only one photometric band is provided \citep{mcgaughm2l}. We discuss the effect of uncertainties arising from our choice of the stellar $M/L$ in \S\ref{sec:uncertainties} and of the dark matter density profile in \S\ref{sec:profile}.
This approach assumes that star formation does not alter the dark matter profile.  In practice, the condensation of gas to the center may lead to dark matter halo contraction \citep{blumenthal}; and vice versa, feedback from stars may lead to expansion/coring \citep{pontzen}.  Besides the fiducial approach of assuming an NFW profile, we also test inferring masses using cored Burkert profiles \citep{burkert_halo}.

We iterate to find the best-fit dark matter profile from within the adopted family of NFW profiles.
We define a trial NFW model by setting M$_h$ and evaluating its concentration parameter using the mean relation between concentration and mass \citep{maccio}. We use GalPy \citep[][http://github.com/jobovy/galpy]{bovy} to evaluate M$_{e,DM}$ and compare to our empirical estimate. We evaluate models over a range of M$_h$  to find the best fit halo. The calculations are done for the adopted cosmology and a redshift of 0.01 to correspond to an overdensity of 346 relative to the matter density \citep[cf.][]{bryan}. For the best fit halo mass, we then add back the baryonic mass using the universal baryon fraction to estimate M$_h$. This is certainly an upper limit to the baryon content and some suggest low mass halos have far less than their 'fair share' of baryons \citep{papastergis}. We will explore the effect of adopting the lower limit in \S\ref{sec:uncertainties}. This approach, including the estimation of $\sigma$ from the scaling relation, was first applied to examine the relation between the number of globular clusters in a galaxy, N$_{GC}$, and M$_h$ \citep{zar22}. The resulting linear relation between N$_{GC}$ and M$_h$ is circumstantial supporting evidence for the accuracy of our estimated M$_h$\ values, modulo the normalisation factor.
    
We ignore scatter in the halo mass-concentration relation, which simulations show is significant
\citep[$>0.1$ dex;][]{maccio}.
\cite{zar22} noted that ignoring the scatter may, for a subtle reason, be the correct approach in this method. The estimates of the internal kinematics of these galaxies is based on the scaling relations, which also sidestep variations among individual galaxies to provide a `typical' $\sigma$ and enclosed mass for each galaxy. Therefore, because the $\sigma$'s we use do not include the effects of differences in the concentration among galaxies of equal mass, the use of mean concentration-halo mass relation may indeed be appropriate.  

The consideration of our treatment of scatter in halo concentration raises a significant concern. How can we verify our estimates of M$_h$? Indeed, \cite{gannon} demonstrated that if cored DM density profiles are adopted, rather than NFW ones, the result can be to invert the relation between M$_e$ and M$_h$. Given that we do not have direct measurements of M$_h$ on a galaxy-by-galaxy basis \citep[even for the Milky Way the M$_h$ estimates show a significant range of values;][]{shen}, we must rely on circumstantial evidence for now. As already mentioned, the resulting linear relation between N$_{GC}$ and M$_h$ is one such piece of evidence. In the case of the SMHM relation (\S \ref{sec:smhm}), bear in mind that either an inverted relation between M$_e$ and M$_h$ or, perhaps more likely, large scatter between M$_{e,DM}$ and M$_h$ --- as would result from including concentration scatter in the DM profiles without accounting for offsetting differences in $r_e$ --- would not lead to the relatively tight SMHM relation we find that closely tracks that obtained using abundance matching techniques.
This is perhaps a less-than-satisfying justification of the approach, but on the other hand offers an avenue for placing constraints on the possible range of variations in the M$_{e,DM}$-M$_h$ relation using the degree of agreement between independent determinations of the SMHM relation.

\section{Results}
\label{sec:results}

\begin{figure*}
\centering
\includegraphics[scale=0.4]{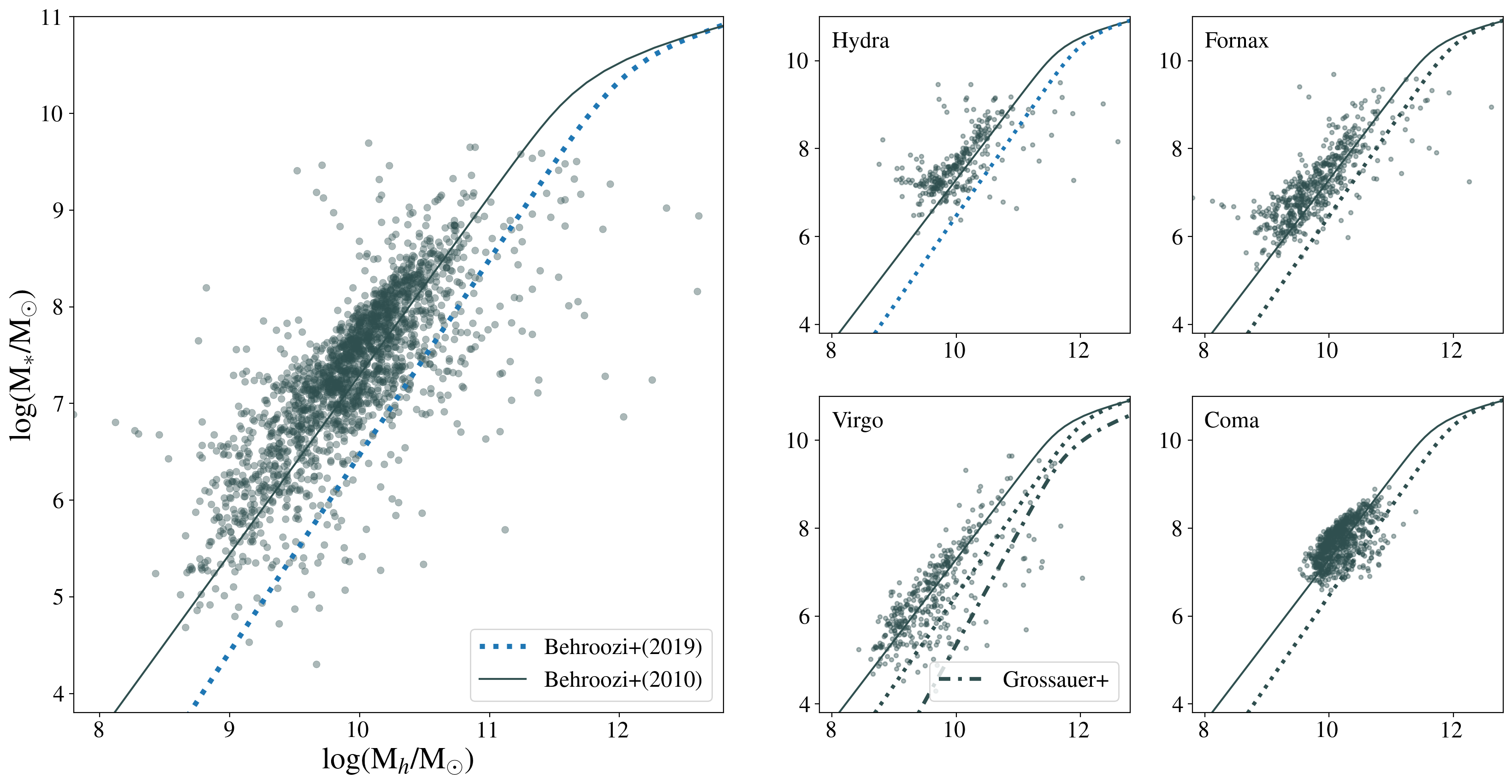}
\caption{The SMHM relation for low-luminosity galaxies in nearby galaxy clusters. Results for each of four clusters (Hydra \citep{lamarca}; Fornax \citep{venhola}; Virgo \citep{virgo}; and Coma \citep{yagi}) are presented on the the right. The composite of these four populations is presented on the left. 
The solid and dotted lines are the extrapolation of the SMHM relations 
from \citet{behroozi} and \citet{behroozi19}, 
respectively, and are the same in all panels. 
The dash-dotted line in the Virgo panel is the SMHM relation from
\citet{Grossauer}, which is particularly interesting because it was derived using the same Virgo galaxy sample, but with a different technique (abundance matching).}
\label{fig:msmh_cluster}
\end{figure*}

\subsection{SMHM relation for local galaxy cluster populations}
\label{sec:smhm}

Large samples of low luminosity galaxies are difficult to obtain because spectroscopy is generally necessary to determine a distance and a luminosity. The standard way to avoid this observational expense is to study low luminosity populations in nearby galaxy clusters, for which one can simply assign the cluster distance to every faint galaxy. There is some danger of background contamination, but the projected density of cluster members at the relevant magnitudes is significantly larger than that of the background and this contrast is even more pronounced for galaxies with low surface brightnesses and relatively  large angular size --- which generally describes the nearby cluster dwarf galaxy population. 

In the application of our methodology to large galaxy samples we are likely to be including all morphological types, unless care is taken to classify and select subsamples. Fortunately, the basic scaling relation we use is applicable to all morphological types \citep{zar08}, so no morphological pre-selection is required. The only distinction in applying the relation to rotation vs. dispersion dominated systems is whether one uses the circular velocity or the velocity dispersion. When using the circular velocity, one needs to divide the value by $\sqrt 2$ 
\citep[the exact value depends on the nature of the potential, the stellar orbits, and radial distribution of stars, but empirical study shows only a weak dependence on this value;][]{weiner,zar08}. However, this distinction is irrelevant for our purposes because at this point we are neither using measured kinematics or estimating the kinematics.

To support this claim, we apply our method to the clean sample of \cite{read17} (avoiding 'rogues' for which they have less confidence in their derived parameters). That study provides all of the necessary information once we convert from their stellar exponential scale radii to effective radii by multiplying their values by 1.68.
For those 9 galaxies, from which they derive total mass using model fitting to H{\small I} rotation curves, our estimates of M$_h$ deviate on average from their quoted M$_{200}$ values by 0.044 dex ($\sim$ 10\%) and have an rms difference of 0.24 dex, a value smaller than what we will eventually find to be observational scatter for our full sample.
We confirm that we can apply our methodology even to H{\small I}-dominated, rotationally-supported low mass galaxies.

There are excellent published catalogues for low luminosity galaxies in the Virgo \citep{virgo}, Hydra \citep{lamarca}, Fornax \citep{venhola}, and Coma \citep{yagi} clusters. The resulting SMHM relations for each of the four clusters, as well as for the composite sample, are presented in Figure \ref{fig:msmh_cluster} and compared to the extrapolations of the \cite{behroozi,behroozi19} SMHM relations. 
The Virgo data trace the relationship to the smallest M$_h$'s among the four samples and the Coma data are the richest, but all are, in the mean, either consistent or only slightly above the \cite{behroozi} curve and consistent with each other. Together, the samples define a clear ridge-line in the M$_*$-M$_h$ space for $9 \lesssim$ log  M$_h/M_{\odot} \lesssim 11$.

One aspect for potential study is highlighted in the panel showing the results for the Virgo galaxies. There we have included the SMHM relation from \cite{Grossauer}, which was derived from the same \cite{virgo} sample of galaxies using an analysis involving abundance matching. Accepting that the technical aspects, such as completeness corrections, were handled properly, the offset between this relation and our results might indicate an anomaly in the halo distribution in the models that were used. The sense of the discrepancy is that \cite{Grossauer} effectively had to place a galaxy with a specific M$_*$ in a more massive halo than that which we are associating it with, suggesting a surfeit of halos in the simulations at these masses. This, in turn, could indicate that halo disruption is underestimated in those models. The general sense of the offset, that abundance matching approaches tend to place galaxies in more massive halos, is consistent with recent considerations of the Milky Way and M 31 \citep{mcgaugh}, although, as we will stress later, the normalisation of our SMHM relation is subject to systematic uncertainties.

Of course, as interesting as such a conclusion might be, it is predicated on the confidence we can place on our overall normalisation of both M$_*$ and M$_h$. The question of M$_*$ can be  addressed by consistently estimating M$_*$ and looking at the situation in a relative sense (in other words, if, for example, the wrong stellar initial mass function is used, as long as the same incorrect assumption is made in both analyses then at least the M$_*$ part of the comparison is valid). The question of M$_h$ can be addressed by spanning a sufficiently large range of M$_h$ that we probe both the power law behaviour at low M$_h$ and the turnover at higher M$_h$. The current difficulty in doing so is that such an analysis requires splicing disparate samples, as we will see below. 

\subsection{The M$_{e,DM}$/M$_{e}$ criteria}
\label{sec:me_criteria}

There are regimes where we might expect our methodology for inferring M$_h$ to perform poorly or not at all. For example, as M$_e$ becomes increasingly dominated by stars our calculation of the dark matter mass within $r_e$, $M_{e,DM}$, will become increasingly uncertain. In fact, errors in our estimate of M$_*$ could even lead to formally negative, unphysical, values of M$_{e,DM}$. As such, we need to reject systems below some value of M$_{e,DM}$/M$_{e}$. This also works to reject the highly compact systems that we excluded in our scaling relation discussion (i.e., those with $r_e < 10$ pc). At the other extreme, systems with extremely high apparent values of M$_{e,DM}$/M$_{e}$ are unlikely to be real because those galaxies would have a baryon fraction far below the universal value. Such systems are most likely due to an underestimation of M$_*$, which leads to an overestimation of M$_{e,DM}$. Because of the large extrapolation from M$_{e}$ to M$_h$, small errors in M$_{e,DM}$ can lead to unphysically large values of M$_h$. As such, we also anticipate needing to set an upper limit on M$_{e,DM}$/M$_{e}$. 

\begin{figure}
\centering
\includegraphics[scale=0.25]{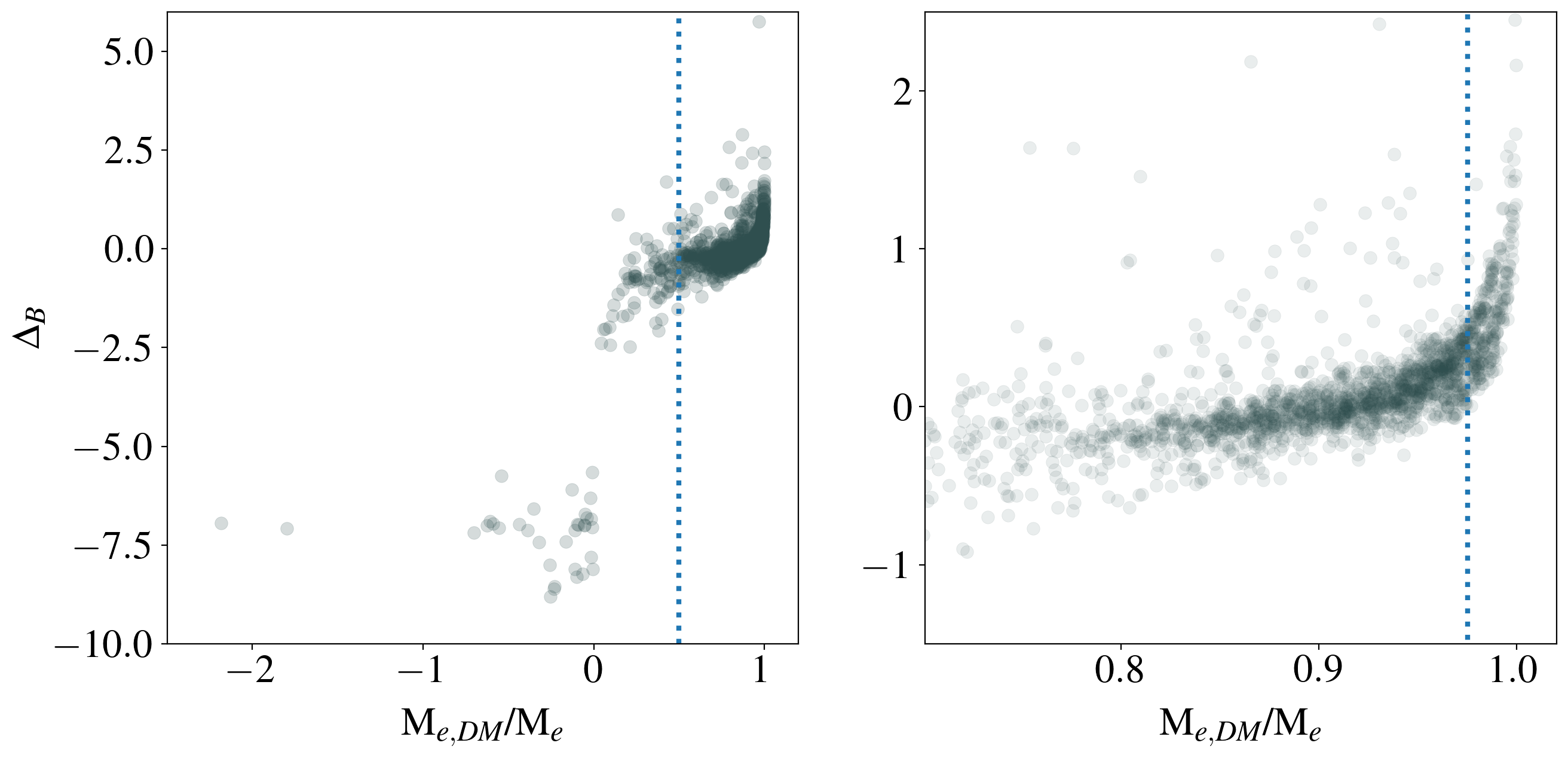}
\caption{Deviations from the \citet{behroozi} fiducial relationship, $\Delta_B$, as a function of the dark matter mass ratio with a sphere of radius $r_e$, M$_{e,DM}$/M$_{e}$, for the set of galaxies shown in Figure \ref{fig:msmh_cluster}. The left panel shows all of the data, while the right one zooms in on the bulk of the data. The dotted lines represent our upper and lower criteria for M$_{e,DM}$/M$_{e}$ going forward.}
\label{fig:mass_ratio}
\end{figure}

To explore these issues we use the results presented in Figure \ref{fig:msmh_cluster} and examine the deviations about the \cite{behroozi} fiducial. We present the deviations from this fiducial, $\Delta_B$, as a function of $M_{e,DM}/$M$_{e}$ in Figure \ref{fig:mass_ratio}. There are a few galaxies with unphysical results (M$_{e,DM}/M_{e} < 0$) because, as anticipated, our estimated value of M$_*$ occasionally exceeds that of M$_{e}$. We reject these cases but they comprise only 1.7\% of the overall sample. Next, we notice that the main distribution in the Figure has an curved shape, with $\Delta_B$ values trailing lower as M$_{e,DM}$ approaches zero. Lower values of $\Delta_B$ correspond to underestimates of M$_h$ relative to the fiducial, which would be expected if scatter moves M$_{e,DM}$/M$_{e}$ below its true value. This downward tail is most visible for M$_{e,DM}$/M$_{e} < 0.5$, so we define a requirement that the ratio exceed 0.5. The dotted line in the left panel of Figure \ref{fig:mass_ratio} shows this cut. At the other end of the M$_{e,DM}$/M$_{e}$ range there is a sharp rise in $\Delta_B$. Here, scatter causes an underestimate of M$_*$, hence an overestimate of M$_{e,DM}$, and a, when extrapolated, a catastrophic overestimate of M$_h$. This sharp rise becomes most prominent for M$_{e,DM}$/M$_{e} > 0.975$, so we set that value as the upper limit. That cut is shown as the dotted line in the right panel of Figure \ref{fig:mass_ratio}. The application of these two criteria removes much of the most egregious scatter from Figure \ref{fig:msmh_cluster}.

Alternatively, a future treatment of this problem could attempt to recover M$_h$ and the associated uncertainty using a Bayesian approach. Our hypothesis is that the recovered values of M$_h$ that result in high $\Delta B$ would also have associated large uncertainties. If they do not, then there is either a tail of systems with intrinsic large scatter in the SMHM relation or a missing ingredient in our model. 

\subsection{How uncertainties affect the results}

\label{sec:uncertainties}

Although the mean trend between M$_*$ and M$_h$ is well-defined in Figure \ref{fig:msmh_cluster}, there is significant scatter about that mean even after we have removed the most egregious outliers using the M$_{e,DM}$/M$_e$ criteria we just described. To better understand the origin of the scatter and how one might lower the observational scatter, we quantify the effects of errors in each of the key parameters in Figure \ref{fig:errors}. For each quantity, we assess the impact by altering the specific parameter by the value shown.

\subsubsection{Distance} 

\begin{figure}
\centering
\includegraphics[scale=0.27]{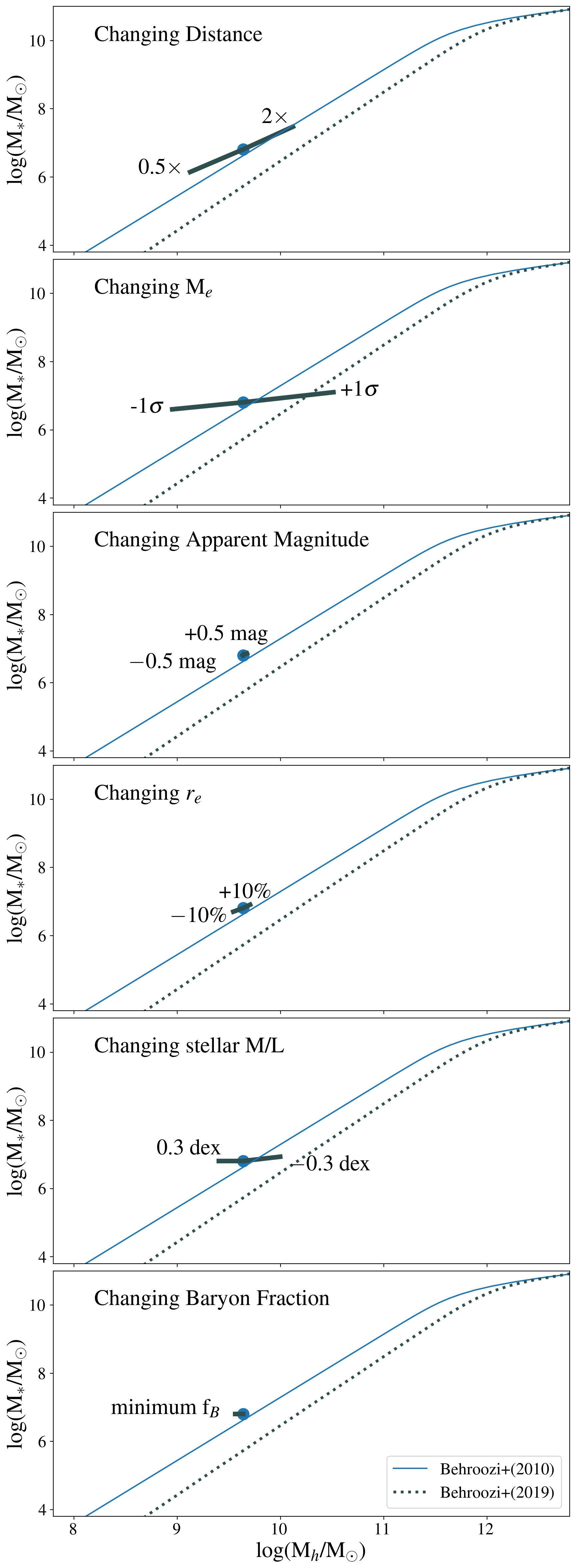}
\caption{The sensitivity of the resulting SMHM relation is shown for the mean location of the Virgo sample \citep{virgo} for a variety of choices. In some cases the range explored matches the plausible uncertainties, in others it does not. Where the range was expanded, it was done to aid visualisation of the effect. }
\label{fig:errors}
\end{figure}

In Figure \ref{fig:errors}, we show how specific changes in one quantity at a time move the mean location of the Virgo sample in the M$_*$-M$_h$ space. The upper panel shows the result of doubling and cutting in half the adopted distance. This is a far larger change than we anticipate, particularly for the cluster galaxies whose hosting clusters are well studied. Distance estimates are more uncertain for individual field galaxies where the peculiar velocities could be significant. Nevertheless, we find that changes in the distance, even when unrealistically large, do not contribute significantly to the scatter because they act to slide sources nearly parallel to the fiducial relation. 

\subsubsection{M$_e$}

We depend on the scaling relation to estimate M$_e$. The scatter in M$_e$, evaluated relative to values obtained for systems with measured $\sigma$'s, is moderate (a factor of 2 in mass) in comparison to the many orders of magnitude in mass over which we apply the relation (Figure \ref{fig:wolf_test}). Even so, those uncertainties are sufficiently large that they can lead to significant errors in the extrapolated estimate of M$_h$.
In the second panel of Figure \ref{fig:errors} we illustrate the effect of a $\pm 1\sigma$ systematic change in the inferred M$_{e}$. Because of the large extrapolation in going from M$_{e}$ to M$_h$, these changes have large repercussions. An initially puzzling aspect of this panel is that a change in M$_{e}$ appears to result in a change in M$_*$, which is an unrelated quantity. This happens because the change in M$_e$ couples to the  M$_{e,DM}/$M$_{e}$ criteria and results in somewhat different samples for which the means are evaluated. 

A second surprising finding is that the observed scatter in Figure \ref{fig:msmh_cluster} is not as large as the result in Figure \ref{fig:errors} would suggest (the scatter about the mean relation in Figure 4 is 0.3 dex while the size of the plotted error bar in each direction is about 0.8 dex). This amplification of the error comes about due to two amplifying effects. First, a more massive halo is also larger and hence $r_e$ is proportionally further inside the halo and, second, the concentration of more massive halos is smaller. These two effects collaborate to turn a 0.1 dex offset in M$_{e}$ into a 0.5 dex offset in M$_h$. 

If the scatter in Figure \ref{fig:wolf_test} comes primarily from scatter in the application of the scaling relation, then
the smaller than expected scatter in the SMHM relation may indicate that the errors in M$_e$ are correlated with a change in another parameter that results in galaxies moving somewhat less across the SMHM relation than indicated in Figure \ref{fig:errors} and more along it. Alternatively, if the scatter in  Figure \ref{fig:wolf_test} comes mostly from scatter in the \cite{wolf} masses, for example due to observational errors in $\sigma$, then the smaller than expected scatter in Figure \ref{fig:msmh_cluster} could be the result of adopting the 'typical' values of M$_e$ given by the scaling relations.

From our analysis we cannot determine the actual, intrinsic scatter in the SMHM relation. While it could be smaller than what we measure, buried underneath the scatter generated by our crude approach, perhaps it is larger than what we are see because we have imposed a degree of homogeneity that does not exist (for example, due to our neglect of scatter in the halo-mass-correlation relationship). Although, measuring the scatter in the SMHM at low masses is a challenge, a value consistent with what we observe is within limits presented elsewhere \citep{allen}, and hence does not point to any catastrophic error in our analysis. Independent derivations of the SMHM scatter would allow us to use our results to provide constraints on possible dark matter density profiles.


\begin{figure}
\centering
\includegraphics[scale=0.27]{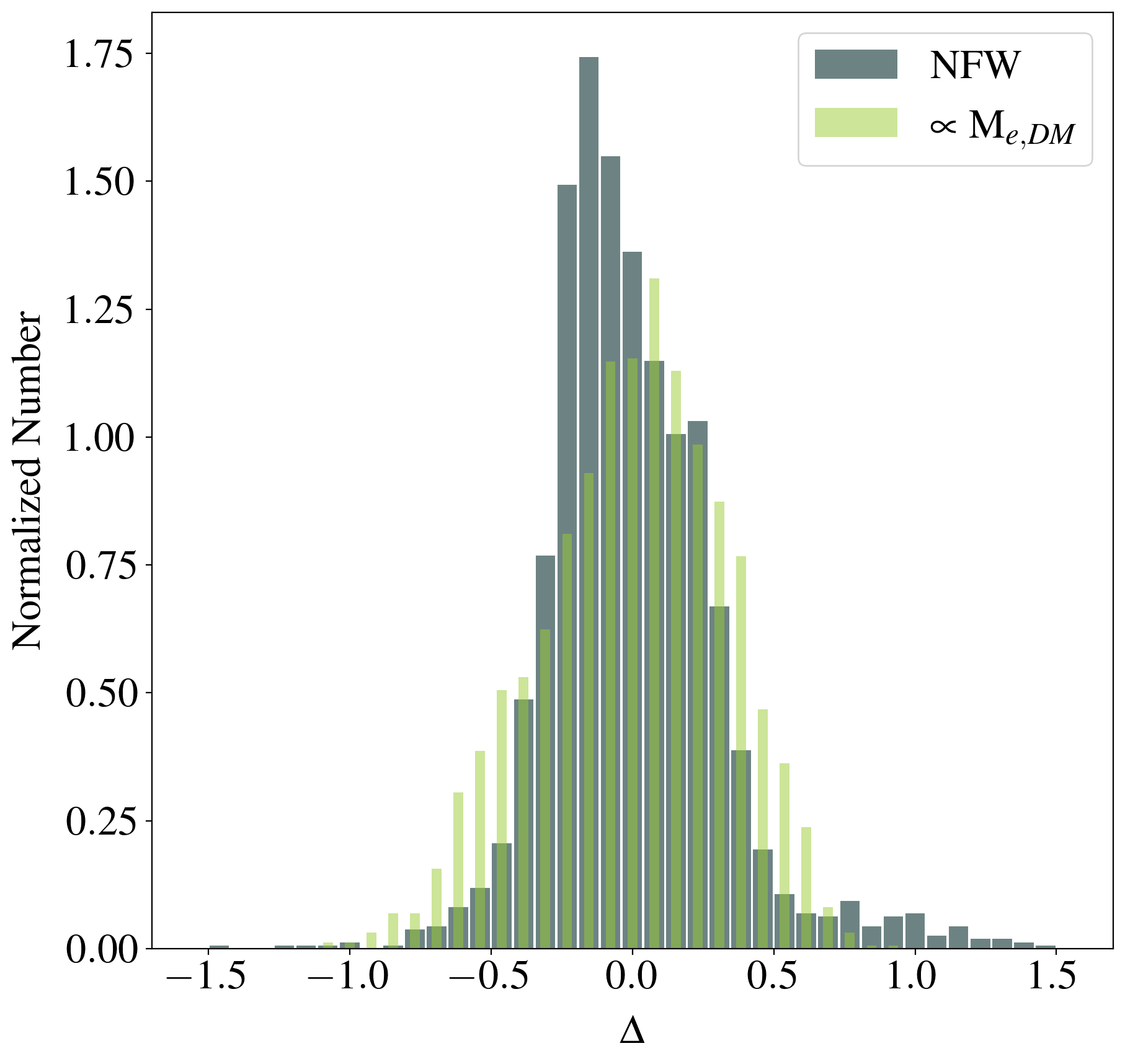}
\caption{A comparison of residuals about the best fit SMHM relations using our approach (NFW; darker, wider bars) a simple scaling of M$_{e,DM}$ ($\propto M_{e,DM}^\alpha$; lighter, narrower bars). The generally narrower distribution of residuals for the standard approach indicate that there is physical information in the extrapolation to M$_h$ using the NFW models. The few outliers ($\Delta > 0.5)$ are again a demonstration of the potential for catastrophic failure in a small fraction of the sample.}
\label{fig:fitcomp}
\end{figure}

\begin{figure*}
\centering
\includegraphics[scale=0.4]{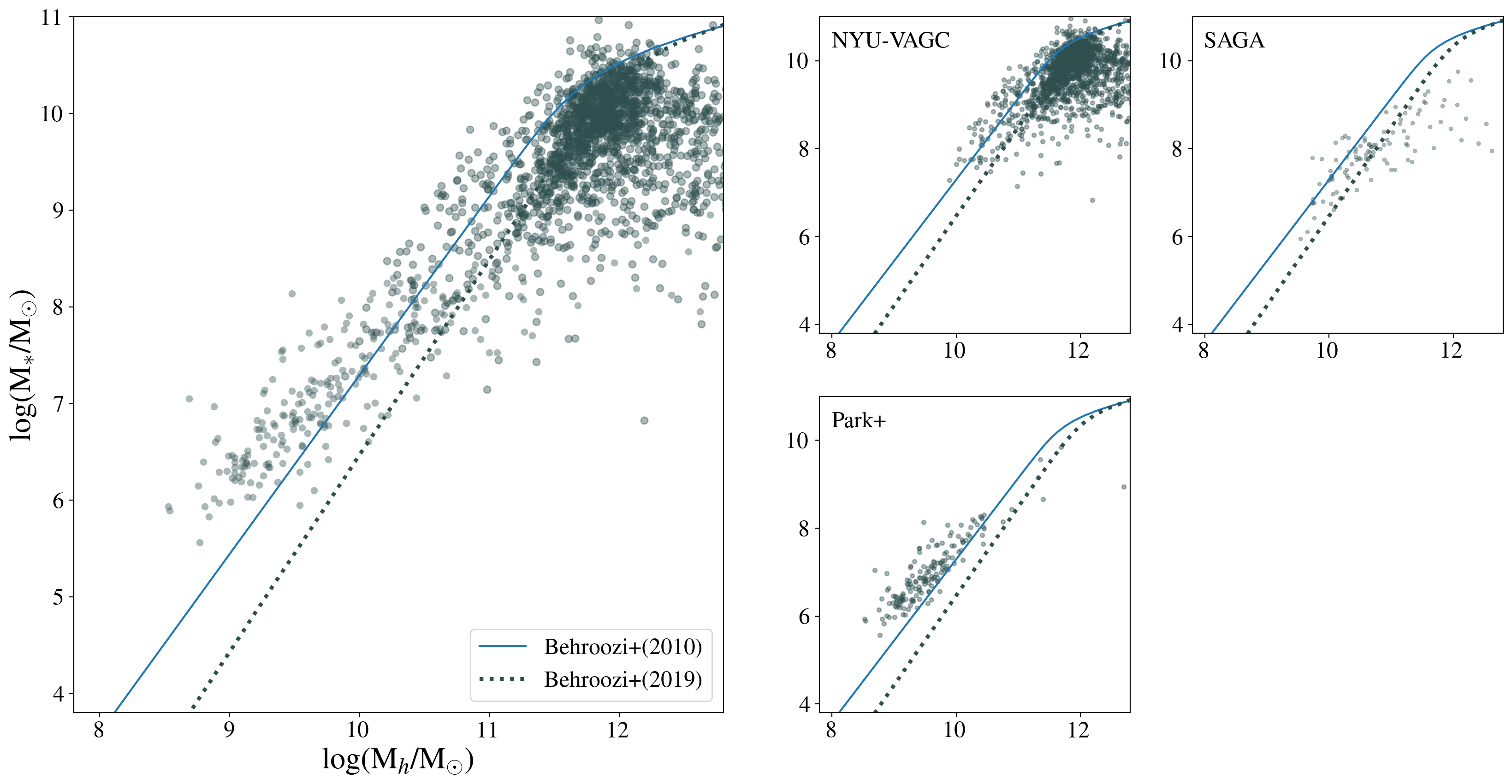}
\caption{The SMHM relation for low-luminosity galaxies in low density environments. Results for two samples probing lower density environments (local volume dwarf galaxies \citep{carlsten21}; satellites of Milky Way analogs \citep[SAGA;][]{geha,mao} are presented on the the right. The SAGA sample has a tail toward very high halo masses that we believe to be spurious. We increase the minimum dark matter fraction requirement to 0.9, the restricted sample, and the tail is mostly removed. 
The composite of the \citet{carlsten21} and restricted SAGA sample is presented on the left.
The solid and dotted lines are the extrapolation of the SMHM relations 
from \citet{behroozi} and \citet{behroozi19}, 
respectively, and are the same in all panels. }
\label{fig:msmh_field}
\end{figure*}

\subsubsection{$I_e$ and $r_e$}

Our determination of $M_e$ depends only on the distance, $I_e$ and $r_e$. We explored the effect of distances errors above and now explore the effects of errors in the other quantities, propagated through the determination of M$_e$. 
In the next two panels of Figure \ref{fig:errors} we show that a much larger than anticipated error in the apparent magnitude and plausible errors in $r_e$ both contribute negligibly to the scatter about the fiducial line. Neither appears to provide enough of a change to help counter the effect of a change in $M_e$. We are left with the conclusion that our estimates of $M_e$ must be somewhat better than reflected in the 0.17 dex scatter in Figure \ref{fig:wolf_test}. Part of the explanation must lie with scatter in the measured $\sigma$'s, which are particularly difficult to measure for low mass systems. A second part may lie with the same hypothesis we made for ignoring scatter in the halo mass-concentration relation. The scaling relation gives an idealised estimate of M$_e$ and is therefore providing an average M$_e$ for similar galaxies, which by the nature of averages has less scatter than that visible in Figure \ref{fig:errors}. As such, we may be in the seemingly absurd regime where having less information (i.e. not having a measured $\sigma$) leads to a more precise result --- as long as the scatter about the scaling relation is proportionally less than the observational scatter in $\sigma$. 

\subsubsection{Stellar M/L}

We now consider two systematic uncertainties that affect the estimation of M$_h$. First, and in the fifth panel in Figure \ref{fig:errors}, we consider plausible changes in the adopted stellar M/L. Here we have adopted a factor of two change downward and upward in the stellar M/L. These changes principally result in a lateral shift in M$_h$, with an amplitude similar to the uncertainty arising from different extrapolations of the SMHM relation. This result highlights the difficulty in using these results to determine the absolute normalisation of the SMHM relation and its dependence on other factors such as the stellar initial mass function.

\subsubsection{Baryon fraction}

To evaluate M$_h$ we assigned each halo a baryon mass determined from the universal baryon fraction. This almost certainly an overestimation of the baryons in each halo \citep{papastergis}, and therefore of M$_h$, although some studies do not find evidence of greater baryon loss in low mass galaxies \citep{geha06}. Nevertheless, to probe the possible full extent of mischaracterising the baryon fraction, we adopt the other extreme of this correction and only add the observed stellar mass to the dark matter halo mass to obtain M$_h$. In the bottom panel of Figure \ref{fig:errors} we show the effect of making that correction instead. Unsurprisingly, the change is visible, but minimal given that a 16\% change in halo mass corresponds to a change in M$_h$ of only 0.075 dex. Because the proper correction must lie between these two alternatives, the effect of adopting the improper correction within these extremes is even smaller and therefore a minor source of uncertainty relative to other issues we have discussed.

\subsubsection{The impact of the adopted potential}
\label{sec:profile}

Comparisons between data and models are also affected by the extrapolation we make from M$_e$ to M$_h$. 
The analysis presented so far is predicated on the adopted NFW dark matter mass profile. However, there is extensive literature advocating alternative profiles \cite{} to resolve some apparent empirical discrepancies between data and the predictions of NFW-based models, particularly among lower mass galaxies \cite[e.g.,][]{burkert20}. Here we briefly discuss the qualitative impact of loosening the adoption of the NFW profile on our results.

Cored potentials offer a larger, and somewhat degenerate, set of models that can fit single radius kinematic constraints \citep{gannon}. In fact, those authors showed that, at least for the UDGs that they were considering, it was possible among some plausible models to invert the relationship between M$_e$ and M$_h$. This raises the important question of whether there is any value in extrapolating measurements of the enclosed mass at small radii to estimates of M$_h$.

The most extreme scenarios, where either the relationship between M$_e$ and M$_h$ is inverted or there is no information in M$_e$ regarding M$_h$, can be rejected on the grounds that we do recover a SMHM relation in qualitative agreement with that recovered from abundance matching studies. 

The more subtle question of whether fitting NFW profiles adds any value, or whether one could simply scale upwards the values of M$_{e,DM}$, requires a quantitative exploration.
For galaxies within our limits on M$_{e,DM}$, we fit for a power law relation between M$_*$ and M$_{e,DM}$. We compare the residuals from that fit, to the residuals from fit for the SMHM using our NFW fitting approach in Figure \ref{fig:fitcomp}. Aside from a sparsely populated tail of large outliers (due to the methods strong sensitivity to M$_{e,DM}$ errors), the results from our NFW fit show less scatter than the uncorrected values. Because the application of the NFW fitting is unlikely to result in a tighter SMHM if there is no relation between M$_e$ and M$_h$, we conclude that our estimates of M$_h$ do add value to the investigation.

\subsection{SMHM for local field populations}

Although cluster dwarf galaxy samples have the advantage of sample size, they have a significant disadvantage in terms of interpretation. Cluster galaxy populations are subject to various effects \cite{gunn,larson,valluri,moore}, and as such may not be representative of the general galaxy population. To address this issue, we examine three sets of field populations of galaxies that include dwarfs \citep{blanton,mao,park1,park2}. The \cite{blanton} sample consists of low z galaxies from SDSS, reanalyzed to improve the photometry for systems of large angular extent, the \cite{mao} sample is from the SAGA spectroscopic survey for satellites of Milky Way analogues \citep{geha-saga}, and the third set comes from an ongoing survey of nearby poor groups  \citep{park1,park2}. The Park et al. sample is different than the other two samples in that distances are assigned from the group membership rather than from recessional velocities. Because their analysis suggests only $\sim$ 30\% contamination and distance errors tend to move galaxies along the SMHM fiducial, we include their sample to  extend coverage down to M$_h \sim 10^8$ M$_\odot$. We convert from Johnson to SDSS photometric bands using the transformations presented by \cite{jester}.

In Figure \ref{fig:msmh_field} we show the results for the three samples, both separately for each sample and together. For the \cite{blanton} sample we have excluded galaxies with $m_r > 18$, which showed far larger scatter than their brighter counterparts, suggesting poor photometry at the faint end of their range. The three samples together cover a large range in M$_h$ and fall between the two plotted extrapolations of the SMHM relations. 

Interestingly, at higher masses they appear to show an offset relative to the \cite{behroozi} fiducial, which the cluster galaxies followed closely (albeit at lower M$_h$), and then fall in line with that fiducial once in a M$_h$ range below $10^{10}$ M$_\odot$.
This behaviour could be a reflection of the fact that dwarf field galaxies tend to be star forming unless they are of very low mass, and because they have not yet formed all of their stars they lie below the fiducial SMHM relation.  This is consistent with findings from several past studies, in which satellite galaxies have larger stellar mass to halo mass ratios than field galaxies \citep[e.g.,][]{rp12,behroozi19}.
It may also simply reflect a shallower SMHM relation than the extrapolated \cite{behroozi} relation.

These are intriguing, although  preliminary interpretations. Comparison across samples is complicated by different measurement techniques, for example the definition of total magnitudes and extinction corrections. Even where we have tried to homogenise the analysis, by correcting to one set of stellar mass-to-light ratios, the correction is often hampered by a lack of similar colour information and photometry in different photometric bands. Even with homogeneous data it will continue to be challenging to obtain absolute values of quantities like the stellar mass-to-light ratio, which depends on the poorly known low end of the stellar mass function. However, if the data are homogeneous and the analysis is done consistently, then relative values will be meaningful and comparisons as that done here can be confidently made.

\section{Discussion}
\label{sec:discussion}

We close by extending the technique to lower mass galaxies in the Local Group. This leads to some mixed results that motivate speculation on the nature of the mass distribution in some of these systems.

\subsection{The composite SMHM relation extended to Local Group galaxies}

\begin{figure*}
\centering
\includegraphics[scale=0.37]{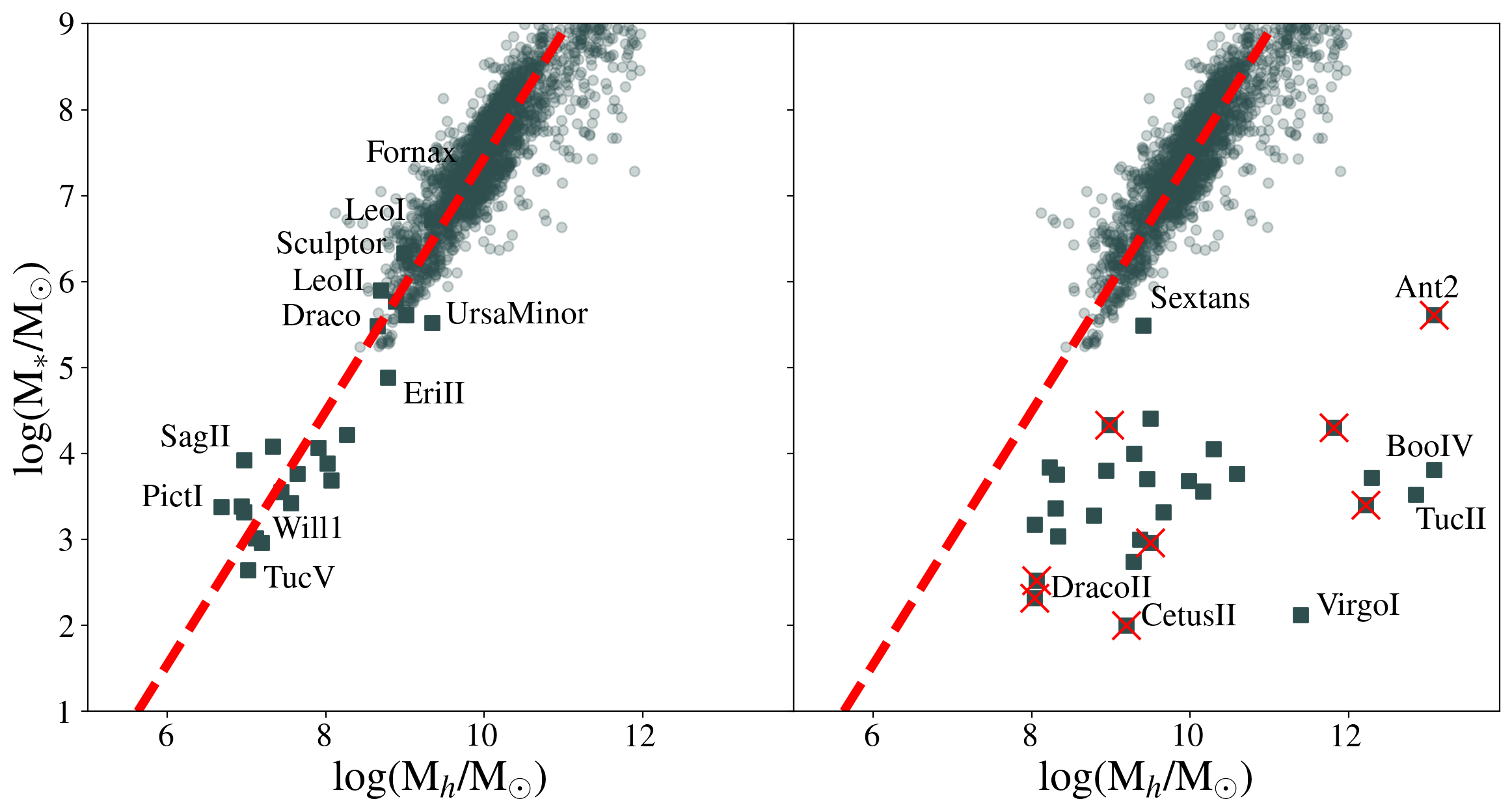}
\caption{The low mass (M$_* < 10^9$M$_\odot$) M$_*$-M$_h$ relation for the combination of the cluster and field subsamples (in lightly coloured circles). Power law fit given in Eq. \ref{eq:smhm} is shown in red dashed lines. The Local Group members from the \citet{nadler} compilation for which we can obtain reliable M$_h$ estimates ($M_{e,DM}/M_e < 0.99$) are shown as squares and labelled in the left panel. Those for which we obtain unreliable mass estimates are shown in the right panel. Some galaxies are labelled to help provide context. The red crosses indicated those for which claims of tidal distortion exist in the literature \citep[][and references therein]{mutlu}.}
\label{fig:composite}
\end{figure*}

In Figure \ref{fig:composite} we present all of the data discussed so far to track the global SMHM relation.
Fitting a power law to those low mass galaxies (M$_* < 10^9$ M$_\odot$ and M$_h < 10^{12}$ M$_\odot$) yields 
$${\rm M}_h = 10^{10.35\pm0.02}\left(\frac{{\rm M}_*}{10^8 {\rm M}_\odot}\right)^{0.63\pm0.02}
\label{eq:smhm}.$$
The data have a standard deviation of 0.31 dex about the line. 
The measured scatter does not depend strongly on our M$_{e,DM}/$M$_{e}$ upper cut. Removing the criteria that M$_{e,DM}/$M$_{e} < 0.975$ increases the scatter for the sample about the best fit relation to 0.37 dex. We nevertheless apply the criteria because the 12\% of sources above this criteria do significantly affect the fitted parameters of the SMHM relation when they are equally weighted in the fit.

In Figure \ref{fig:composite} we also add our estimates of M$_*$ and M$_h$ for Local Group (LG) dwarf galaxies derived from the data provided in the \cite{nadler} compilation. From that list, we exclude Crater II, which is a challenging galaxy to model in any regard \citep{crater} and Kim 2, Triangulum II, and DES J0225+0304, which have $r_e \sim 10$ pc. This compilation provides $V$ band photometry and we adopt a standard stellar M/L = 1.2 for the remaining 54 galaxies as suggested by \cite{mcgaughm2l} when colours are not available to estimate the stellar mass. We adjust the M$_{e,DM}/M_e$ limit upward to 0.99, to include more galaxies and because these galaxies are generally more dark matter dominated than the more massive galaxies we discussed previously. After applying the new $M_{e,DM}$/M$_e$ criteria, we are left with 24 galaxies and they fall tightly along an extension of the SMHM relation obtained from the cluster and field samples (Figure \ref{fig:composite}, left panel). These galaxies include six of the classic dwarf Spheroidals, with Sextans, the one that does not satisfy the criterion, lying just slightly farther off the mean trend. As such, these systems fall nicely along the extrapolation of the SMHM and, therefore, consistent with model expectations, as found previously to be the case for these galaxies in an independent analysis \citep{read19}.

However, slightly less than half of the LG galaxies survived the  M$_{e,DM}/M_e$ criteria, and those that do not populate the lower right of the right panel of Figure \ref{fig:composite}. This includes some with an estimated M$_h$ that differs from that inferred from their stellar mass by several orders of magnitude. The discrepancy is sufficiently large that simple observational errors cannot be responsible. A natural suspicion falls on the estimated $\sigma$, and related M$_e$, obtained using the scaling relation. However, for the 13 LG galaxies that fall more than 1 dex away from the extrapolated SMHM relation for which we have found a spectroscopically measured $\sigma$ in the literature (Table \ref{tab:sig_comp}), only one has an estimated velocity dispersion that exceeds the measured one by more than 3$\sigma$. 
Although it is worth investigating why some have statistically large deviations, errors in our estimation of $\sigma$ are not responsible for the bulk of the large offsets from the SMHM relation. Nevertheless, given the small values of $\sigma$ for this set of objects there is a concern that large deviations can only work to overestimate $\sigma$, creating a bias in the outliers in one direction.

\begin{table}
\centering
\caption{Comparing $\sigma$ determinations for SMHM LG outliers}
\begin{tabular}{ lrrr} \hline  
Galaxy & Estimate & Observed & Reference \\ 
&(km s$^{-1}$)&(km s$^{-1}$)\\
\hline
Bootes I&7.2&5.1$^{+0.8}_{-0.7}$&\cite{jenkins}\\
Bootes II&3.4&10.5$\pm$7.4&\cite{koch}\\
Coma Berenices&4.4&4.6$\pm$0.8&\cite{simongeha}\\
Grus I&1.9&$2.5^{1.3}_{-0.8}$&\cite{chiti}\\
Grus II & 6.7 & $<$1.9 & \cite{simon2020}\\
Hercules&6.2&5.1$\pm0.9$&\cite{simongeha}\\
Leo IV&6.2&3.4$^{1.3}_{-0.9}$&\cite{jenkins}\\
Segue 1&2.7&3.7$^{+1.4}_{-1.1}$&\cite{simon11}\\
Sextans&8.5&8.9$\pm$0.4&\cite{walker06}\\
Tucana II &10.6&4.6$\pm1.5$&\cite{chiti21}\\
       &10.6&8.6$^{4.4}_{-2.7}$&\cite{walker16}\\
       &10.6&6.2$^{+1.6}_{-1.3}$&\cite{taibi}\\
Tucana IV & 9.3 & $4.3^{1.7}_{-1.0}$ & \cite{simon2020}\\
Ursa Major I&7.9&7.6$\pm$1&\cite{simongeha}\\
Ursa Major II&6.0&6.7$\pm$1.4&\cite{simongeha}\\
\label{tab:sig_comp}
\end{tabular}
\end{table}

The outliers can be interpreted in a variety of ways. First, these may be tidally distorted systems for which the assumption of equilibrium is inappropriate. Given the large number of such systems, this seems unlikely as a blanket explanation, but is likely to be an important factor in a number of cases. For example, Tuc IV appears to have collided with the LMC a mere 120 Myrs ago \citep{simon2020}, Cetus II is an enhancement along the Sagittarius stream \citep{conn}, Tucana III has long tidal streams emanating from it \citep{shipp}, and Draco II and Antlia 2 are believed to be disrupting \citep{longeard,ji}. We have highlighted with red crosses in the Figure those systems for which tidal distortions have been empirically claimed \citep[][and references therein]{mutlu}. Because theoretical modeling suggests that all of these systems should have suffered significantly as a result of tidal interactions \citep{fattahi}, explanations along these lines cannot be easily dismissed.

Second,
the mass estimates could at least be roughly correct, in that these may be systems with unusually low values of stellar mass for their halo mass. Such systems would be examples of relatively massive subhalos with vastly underproduced stellar populations and examples of a large scatter in the SMHM relation at low masses. We disfavour this as a blanket interpretation as well because at the higher end of the mass range (M$_h > 10^{11}$ M$_\odot$) such systems would have macroscopic dynamical consequences on the LG. Furthermore, from the right panel of Figure \ref{fig:composite}, the distribution of outliers suggest a progression to higher M$_h$ at fixed M$_*$ rather than one of lower M$_*$ at fixed M$_h$. 

Lastly, the discrepancy may hint at deviations from our standard dark matter model. 
In our particular problem, the nature of this excess mass is unspecified, and so we explore the possibility that it is in the form of a central black hole. However, these system could also be strong outliers from the mass-concentration relation that we use.

\begin{figure}
\centering
\includegraphics[scale=0.27]{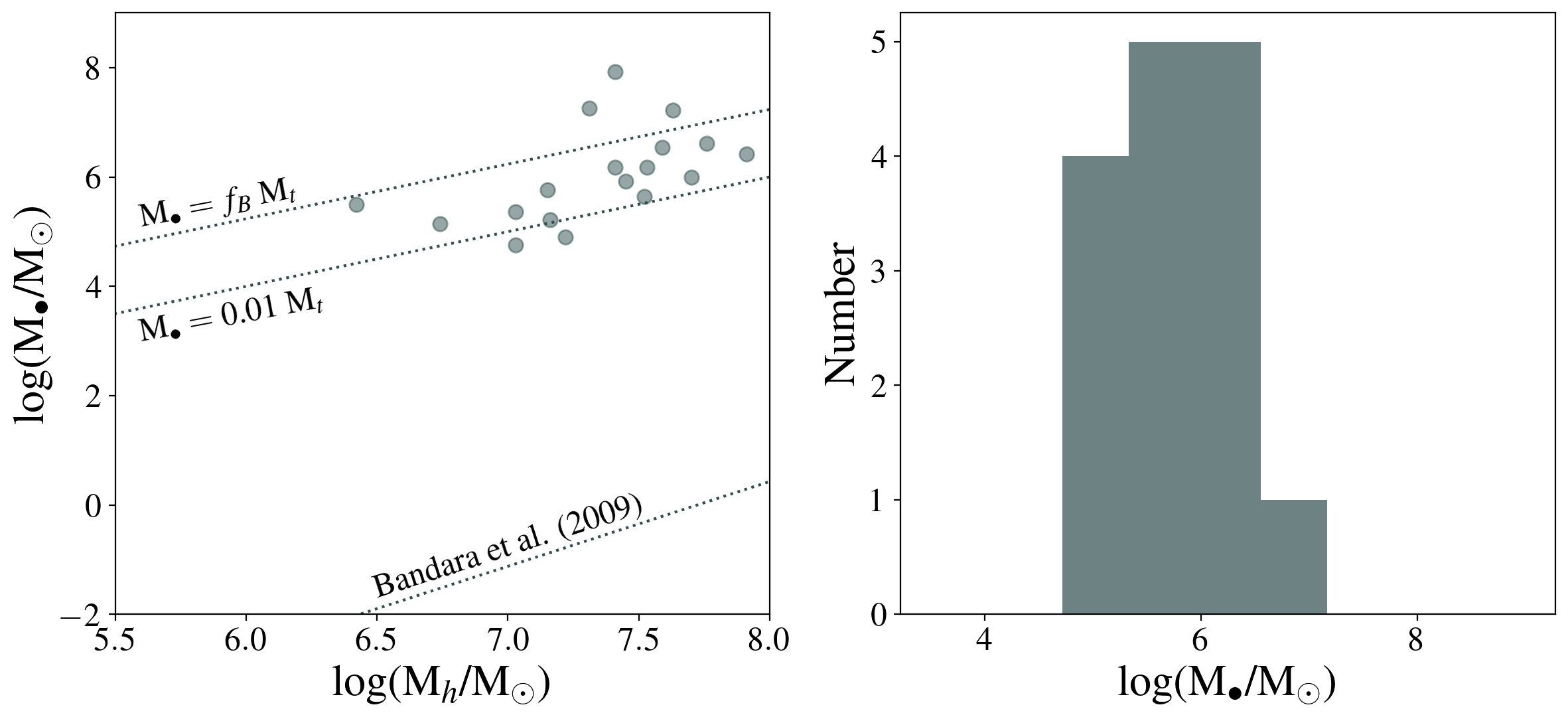}
\caption{Central massive black hole scenario. For those LG galaxies that fall far from the SMHM relation and do not have published claims of tidal distortion, we redo the analysis assuming that they fall on the relation and recover the required central black hole mass to make this happen. That calculated M$_{\bullet}$ is plotted vs. M$_h$ in the left panel. For comparison the extrapolation of the corresponding relation for massive galaxies \citep{bandara},
a line representing M$_{\bullet} = $M$_{h}/100$, and a line assuming all of the baryons in the halo are in the black hole are also shown. In the right panel we show the distribution of recovered M$_{\bullet}$ for the physically plausible cases where M$_{\bullet} < f_B$M$_h$}. 
\label{fig:bh}
\end{figure}

A massive central black hole could contribute a significant fraction of the mass measured within $r_e$ and, therefore, removing that mass from what the standard (e.g., NFW) dark matter halo has to match within $r_e$ will significantly lower the derived halo mass. We now estimate the central black hole masses, $M_\bullet$, needed to place these systems on the extrapolation of the SMHM relation shown in Figure \ref{fig:composite}. We simply refit our model, now subtracting both M$_\bullet$ and 0.5*M$_*$ from M$_e$ to obtain M$_{e,DM}$, and ask what value of M$_\bullet$ places the galaxy nearest the SMHM relation. We do this for all of the galaxies that are at least 1 dex away from the relation in our original analysis and do not already have published claims of being tidally distorted. We can place all of the systems back on the relation and the inferred black hole masses are shown in Figure \ref{fig:bh}. 

The resulting black hole masses are orders of magnitude larger than one might invoke using an extrapolation of the M$_\bullet$-M$_h$ relation for larger galaxies \citep{bandara}, but relations such as this are expected to flatten at low masses \citep{greene}. This model has various implications for black hole seed masses and the evolution of the black hole mass function with time that we do not explore here, but it offers a straightforward way of addressing our difficulty in fitting these systems without invoking exotic dark matter physics. A significant challenge that this scenario faces is that the black hole mass accounts for a large fraction of the baryons expected within these halos (apparently surpassing it in 5 cases), although a combination of observational errors and relaxing the requirement that the galaxies lie exactly on the mean SMHM relation may address the most extreme cases. 
If we reject those three systems for which M$_\bullet > f_B$M$_h$ as being physically implausible, the remainder of the set have $\log$(M$_\bullet/$M$_\odot) = 5.7\pm0.6$ (right panel of Figure \ref{fig:bh}). 

The large inferred values of M$_\bullet$ may appear, and may ultimately prove to be, problematic for this hypothesis. Nevertheless, 
for some systems that fall off the SMHM relation, there is additional information in the literature that we can use to gain  intuition into the relevant uncertainties and test our inferences. Tuc II, which is one of the five galaxies for which the inferred M$_{\bullet}$ is larger than $f_b$M$_h$, and is therefore suspect, is one galaxy for which additional spectroscopic data and modelling of the enclosed mass out to large radii exist and extend well beyond $r_e$ \citep{chiti}. That study provides an estimate of the enclosed mass within 1.1 kpc (2.14$^{+3.67}_{-1.24} \times 10^7$ M$_\odot$), which is $\sim$ 39 times smaller than what we derive ($8.4 \times 10^8$ M$_\odot$) from our baseline model, i.e., one without a central black hole. Of course an overestimate of the mass enclosed at 1.1 kpc leads to an overestimate of the halo mass, which is what led to Tuc II falling so far off the SMHM relation. Our suggested solution of including a central black hole lowers both the inferred halo mass and the inferred mass at 1.1 kpc. For our inferred M$_{\bullet}$, the resulting enclosed mass (halo + BH + stars) at 1.1 kpc drops to $2.4\times 10^7$ M$_\odot$, in excellent agreement with the measurement by \cite{chiti} thereby providing supporting evidence for our suggestion. For completeness, we note that \cite{chiti} adopted $r_e = 120$ pc, as opposed to the 165 pc in \cite{nadler}. Doing the calculation for this different value of $r_e$, our estimate for $\sigma$ drops from 10.8 to 8.4 km s$^{-1}$, in better agreement with \cite{walker16}, the inferred black hole mass drops to 10$^{6.7}$ from 10$^{7.3}$ M$_\odot$ but it is still among the largest of our set and remains close to $f_b$M$_h$, and the 
enclosed mass drops to $1.3 \times 10^{7}$ M$_\odot$, still within the uncertainty range of the measurement by \cite{chiti}. Loosening the criterion that Tuc II lies exactly on the SMHM reduces the inferred M$_{\bullet}$ and increases the enclosed mass at 1.1 kpc, both of which would align even better with expectations. We close by noting that the need for excess central mass in Tuc II remains if one adopts the smallest observed value of $\sigma$ rather than our inferred value.

Although a massive central black hole is one way to address the outliers, this approach for resolving the discrepancies only requires a highly concentrated secondary mass component. This mass component could be a black hole, but it could also be a more tightly bound secondary dark matter component that contributes mass primarily at radii within $r_e$. We do not find that the deviations from the SMHM relation depend on $r_e$, which one might expect if this second component dominates within a physical radius that comparable to $r_e$.

We close by noting that among these alternatives, the only one we know must play a role in at least some of these systems is that of tidal deformation/destruction. As such, it is not necessary for any of the alternatives to be true in every one of the discrepant galaxies. There may be some with a significant error in our estimate of M$_e$, some with lower than expected M$_*$, and some with a nuclear black hole.  At the very least, this discussion highlights which Local Group galaxies merit further attention.

\section{Summary}

We present a photometric halo mass estimation technique for local galaxies. The technique is predicated on 1) the university applicability of the \cite{wolf} mass estimator, 2) our empirical fit to the effective mass-to-light ratio within the effective radius, $\Upsilon_e$, that is second order in $\log \sigma$ and $\log I_e$, where $I_e$ is the mean surface brightness within $r_e$, and 3) the adoption of a dark matter density profile that is used to extrapolate to a halo mass. Each of these has the potential for systematic errors. The first was established using numerical simulations to be valid for spheroidal galaxies \citep{wolf}. Bootstrapping to the universality of the scaling relation presented by \cite{zar08}, the mass estimation should be independent of morphological type. The second we validate by comparing the our resulting estimated enclosed masses within $r_e$ with those obtained with the \cite{wolf} estimator for a sample of galaxies with available measurements of $\sigma$. The last is the most difficult to verify as there are few measurements of the halo mass for individual galaxies. The general behaviour of our mass estimates is indirectly validated by the resulting linear relationship between the number of globular clusters and halo mass when using this methodology \citep{zar22} and by the agreement shown here in the recovered SMHS relation with that extrapolated from abundance matching techniques \citep[e.g.,][]{behroozi}.

We find no detectable difference among the SMHM relations of four local clusters or between the cluster and field relations. We find no change in the slope of the relation for $9 < \log$ M$_h$/M$_\odot < 11$, although the slope across the full mass range explored ($9 < \log$ M$_h$/M$_\odot < 12$ may be shallower than that extrapolated from abundance matching \citep{behroozi,behroozi19}. We fit a power law to our empirical SMHM relation and find that for adopted NFW dark matter profiles and for M$_* < 10^9$ M$_\odot$,
$${\rm M}_h = 10^{10.35\pm0.02}\left(\frac{{\rm M}_*}{10^8 {\rm M}_\odot}\right)^{0.63\pm0.02}.$$ The normalisation is susceptible to systematic errors that depend on the adopted dark matter potential. The slope will have systematic errors if typical dark matter profiles systematically depend on mass.  For example, if dwarf galaxies were more likely to have feedback-driven cores than more massive galaxies \citep{pontzen}, then the slope would be shallower than our fiducial result here, similar to the extrapolated abundance matching results.  We note also that the quoted uncertainties above refer to the uncertainties in the median relation and do not capture the galaxy-to-galaxy scatter. For galaxies with M$_h < 10^{11}$ M$_\odot$ the scatter about the fit in $M_h$ is 0.3 dex inclusive of the uncertainties in our method but with the additional sample cuts described in the \S\ref{sec:me_criteria}.

Finally, we place lower luminosity Local Group galaxies on the relationship using the same technique and find that about half lie well along the extrapolated relationship, but that those with extremely high inferred ratios of dark matter to luminous matter within $r_e$, which we generally rejected in our technique as being unphysical, fall far from the SMHM relationship. If one accepts these values, then the nature of discrepancy is that there is too much dark mass within $r_e$. When we posit that these galaxies indeed do lie on the SMHM and that the extra dark matter mass within $r_e$ does not belong to the larger dark matter component, we can calculate how much extra mass there is. Hypothesising that this mass is in the form of a central black hole mostly yields black hole masses in the range of intermediate mass black hole,  $10^{5.7\pm0.6}$ M$_{\odot}$, and roughly one to a few percent of M$_h$. At the very least, this analysis highlights several Local Group galaxies that merit a closer look.

Our technique provides an independent way to derive SMHM relationships for local galaxy samples. Its power is mostly in enabling statistical comparisons, although it can be used to highlight interesting cases worthy of follow up study, such as in the case of the inferred IMBHs in certain Local Group dwarf galaxies. The empirical basis for the relation means that refinements will be made as the calibrating samples grow in size and provide greater representation of galaxies at the extremes, such as ultra-diffuse galaxies and ultra-compact dwarfs. Nevertheless, it currently provides a valuable independent comparison to the dominant abundance matching approach and provides support for the power-law extrapolation of those results to lower halo masses.

\section*{acknowledgments}
DZ acknowledges financial support from  AST-2006785.  PB was partially funded by a Packard Fellowship, Grant \#2019-69646.


\section*{Data Availability}
No new data were generated or analysed in support of this research.



\bibliographystyle{mnras}
\bibliography{zaritsky.bib} 

\bsp
\label{lastpage}
\end{document}